\documentclass{rnaastex}

\begin{document}

\title{The $^{87}$Rubidium Atomic Clock Maser in Giant Stars}

\correspondingauthor{Jeremy Darling}
\email{jeremy.darling@colorado.edu}

\author[0000-0003-2511-2060]{Jeremy Darling}
\affiliation{Center for Astrophysics and Space Astronomy \\
Department of Astrophysical and Planetary Sciences \\
University of Colorado, 389 UCB \\
Boulder, CO 80309-0389, USA}


\maketitle


\vspace{30pt}

We conducted a Green Bank Telescope (GBT) search for the ground state 6.8 GHz hyperfine transition of 
rubidium ($^{87}$Rb) toward giant stars detected in \ion{Rb}{1} optical resonance lines.  
The spin-flip transition of $^{87}$Rb is one of the principal 
transitions used in atomic clocks, in addition to the 
hydrogen 21 cm maser and the $^{133}$Cs hyperfine transition (which defines the second).  
The optical lines of $^{87}$Rb and $^{85}$Rb can together pump the 6.8 GHz transition to form a maser, and
the same optical pumping used in atomic clocks
may occur in the atmospheres of evolved stars.  

Rubidium ($Z=37$) follows the hydrogen isosequence with valence electron ground state $5\,^2S_{1/2}$.  
$^{87}$Rb has a nuclear spin of $I = 3/2$, so the total angular momentum of 
the ground state including hyperfine interactions can be $F=1$ or $F=2$.  The 
hyperfine transition between these two states has a frequency of 
6.83468261090429(9) GHz \citep{bize1999}, and, with the cesium hyperfine transition, 
is one of the two main atomic clock transitions in general use.  
The primary optical resonance transitions are $5\,^2S_{1/2}\rightarrow5\,^2P_{1/2}$ 
and $5\,^2S_{1/2}\rightarrow5\,^2P_{3/2}$ at 795 and 780 nm, respectively\footnote{\url{http://steck.us/alkalidata/}}.

Rubidium has two common isotopes, $^{85}$Rb (stable) and $^{87}$Rb (50 Gyr half-life), with
terrestrial ratio 72:28 \citep{pringle2017}.  
Rb atomic clocks 
employ the hyperfine structure of the  $^{85}$Rb 
optical resonance lines to selectively filter and thus optically pump the $^{87}$Rb hyperfine ground states, 
creating a population inversion for 6.8 GHz $^{87}$Rb maser action \citep{bender1958,davidovits1966}.
The same physics that are used in Rb atomic clocks may obtain in stellar and sub-stellar
atmospheres, producing a natural 6.8 GHz $^{87}$Rb atomic clock maser.

Astrophysical maser action requires three ingredients:  population inversion
of a metastable state, seed photons, and a velocity-coherent amplification pathway.
Asymptotic giant branch (AGB) stellar atmospheres generally satisfy these conditions for multiple species
and are prodigious emitters of molecular masers such as SiO, OH, and H$_2$O.  
We predict that the conditions in AGB stellar atmospheres and circumstellar shells are promising 
for $^{87}$Rb maser action, which requires similar conditions to those required
for molecular masing.  The stellar targets observed for this study are known to 
host molecular masers.  Moreover, the \ion{Rb}{1} optical pumping lines
have been observed in AGB stars \citep[e.g.,][]{garcia-hernandez2006}, and both isotopes of \ion{Rb}{1} are 
inferred to be present based on the presence of s-process elements such as Zr.

Detection and subsequent development of the $^{87}$Rb maser would provide a 
useful cosmic clock.  By tying clocks in space to terrestrial standards, one can 
test basic physics using telescopes 
and, with {\it Gaia} proper motions, obtain precise three-dimensional kinematics of stars.
If the Rb atomic clock maser can be detected from stars, brown dwarfs, or 
exoplanetary atmospheres, then the line can be referenced to the ground standard
and enable new exceptionally precise measurements.  
 

We searched for the 6.8 GHz $^{87}$Rb maser in 
objects where \ion{Rb}{1} is already detected by optical spectroscopy:  
AGB OH/IR stars that show super-solar [Rb/Fe] abundances \citep{garcia-hernandez2006};
the Mira variable R And -- a SiO and H$_2$O maser-emitting star \citep{kim2010} --- 
showing the 780 nm \ion{Rb}{1} line in emission \citep{wallerstein1992};
and giant stars in the globular clusters M4 and M5 \citep{yong2008}.  
From the OH/IR star sample, we selected stars with [Rb/Fe] at least ten times solar and $\delta > -25^\circ$.  
Many of these stars show \ion{Rb}{1} lines from the stellar photosphere as well as from expanding 
circumstellar shells.  
The globular clusters M4 and M5 show little star-to-star variation in Rb abundance, so we integrated on the 
central 1.8\arcmin.

GBT position-switched observations with 2.5 minute cadence and 187.5 MHz total bandwidth were polarization-averaged and
smoothed to 1.0 km s$^{-1}$ channels.
Table 1 lists the total integration times and rms noise for ten molecular maser-emitting giant stars and two globular clusters.  
No 6.8 GHz $^{87}$Rb lines were detected above 3.8$\sigma$.  
The 6.7 GHz methanol maser was also not detected.

\begin{deluxetable}{crrrc}[t]
\tablecaption{Observations and Results\label{tab:1}}
\tablehead{
\colhead{Object} & \colhead{Coordinates} & \colhead{Velocity\tablenotemark{a}} & \colhead{$t_{int}$} & \colhead{rms\tablenotemark{b}}\\
\colhead{} & \colhead{(J2000)} & \colhead{(km s$^{-1}$)} & \colhead{(s)} & \colhead{(mJy)} 
}
\startdata
R And                   & 00:24:01.68 +38:34:32.9 & 0.0 & 446 & 3.0 \\
IRAS01085+3022  & 01:11:16.12 +30:38:03.7 & 0.0 & 445 & 3.0 \\
IRAS05151+6312  & 05:19:51.81 +63:15:51.9 & 0.0 & 445 & 2.9 \\
IRAS06300+6058  & 06:34:33.69 +60:56:25.2 & 0.0 & 742 & 2.3 \\
M5                        & 15:18:33.75 +02:04:57.6  & 51.8 & 1928 & 1.4 \\
IRAS15576$-$1212  & 16:00:23.76 $-$12:20:57.5 & $-$18.0 & 446 & 3.2 \\ 
M4                        & 16:23:35.40 $-$26:31:31.8 & 70.4 & 1931 & 1.7 \\ 
IRAS18429$-$1721  & 18:45:51.29 $-$17:17:59.6 & 0.0 & 446 & 4.6 \\
IRAS19059$-$2219  & 19:08:54.75 $-$22:14:21.0 & 6.0 & 446 & 80.8\tablenotemark{c} \\
IRAS19426+4342  & 19:44:16.51 +43:49:20.2 & 0.0 & 298 & 3.5 \\
IRAS20052+0554  & 20:07:42.64 +06:03:11.0 & 0.0 & 446 & 3.0 \\
IRAS20077$-$0625  & 20:10:27.90 $-$06:16:12.0 & $-$32.0 & 446 & 3.0 \\
\enddata
\tablenotetext{a}{LSR, optical definition.}
\tablenotetext{b}{Noise per 1.0~km~s$^{-1}$ channel.}
\tablenotetext{c}{Solar interference.}
\end{deluxetable}  

\acknowledgments

\facilities{GBT}

\software{GBTIDL}

\end{document}